\documentclass{article}
\usepackage[dvips]{graphicx}
\usepackage{epsf}
\begin{document}
\baselineskip7mm
\title{Braneworld dynamics with vacuum polarization}
\author{A. V. Toporensky$^{\dagger}$ and P.V. Tretyakov$^{\ddagger}$}
\date{}   
\maketitle
\hspace{8mm}{\em Sternberg Astronomical Institute,  
Universitetsky prospekt, 13,  Moscow 119899, Russia}

\begin{abstract}
We investigate the cosmological dynamics of a brane Universe
when quantum corrections from vacuum polarization are taken into account.
New vacuum de Sitter points existing on Randall-Sundrum brane are described.
We show also  that quantum correction can destroy the DGP de Sitter solution
on induced gravity brane.  
\end{abstract}
$^{\dagger}$ Electronic mail: lesha@sai.msu.ru\\
$^{\ddagger}$ Electronic mail: tpv@xray.sai.msu.ru\\

Investigations of quantum effects in a strong gravitational field 
 and their applications in cosmology  have
a long story. Since the beginning of 70-th many interesting results
modifying the standard Friedmann cosmology due to vacuum polarization
and particle production have been obtained. In particular, a vacuum
de Sitter solution and an inflationary regime
driven solely by vacuum polarization without any matter \cite{SI} was
described even earlier than a common scalar field inflationary
scenario. A detailed analysis of cosmology with vacuum
polarization (we will consider only this effect
in the present paper) have been done in \cite{Hartle}. After some
period of stagnation, this problem begins to attract a considerable
attention last several years, mainly due to development of
modified gravity models. The form of extra terms in cosmological
equations of motion caused by vacuum polarization does not depend
on a particular theory of gravity, however, peculiarities of a
background metric in these theories could result in some new
dynamical regimes.
Recently modifications caused by the vacuum polarization
have been studied for regimes with
soft future singularities. Such regimes, being impossible  in the
standard cosmology, are rather typical in some modern cosmological
scenarios (see for example, \cite{Barrow}), in particular, they
are present in induced gravity brane models \cite{Sht-Sing}. Quantum
corrections change the dynamics significantly, leading to a softer
singularity or even to non-singular solutions \cite{Odintsov, O-Shinji,
we}.

Another interesting problem is stability of classical solutions
with respect to quantum corrections. As quantum terms contain
higher time derivatives in comparison with corresponding classical
equation of motion, some classical solutions may become unstable.
In \cite{we} this instability is described for certain regimes in
induced gravity brane cosmology.

It is well known that the vacuum polarization leads to the following
vacuum expectation value for the energy density:
\begin{equation}
\rho_q=<T_{00}>=k_2 H^4 + k_3(2 \ddot H H + 6 \dot H H^2 -\dot H^2),
\end{equation}
where $H$ is the Hubble parameter,
$k_2, k_3$ depend upon the spin weight of the different fields 
contributing to the vacuum polarization.

If the classical Friedmann equation has the form of {\it algebraic} dependence
of the Hubble parameter upon the energy density of the Universe
$H=H(\rho)$, substituting $\rho \to \rho+\rho_q$ we get a {\it differential}
equation which governs the cosmological evolution with quantum corrections.

It is clear that $k_2$-term in $\rho_q$ can be incorporated into the function
$H(\rho)$. Only $k_3$-term containing time derivatives may provide
instability. In the paper \cite{we}  the case of $k_2=0$ have
been studied in braneworld models. 
It is a good assumption near a soft future singularity
where  $\dot H \to \infty$ while $H$ is finite, and $k_3$-term in (1)
dominates. In the present paper we describe general features of brane
dynamics with quantum corrections in their full form.

We start with modification of classical equations
due to $k_2$-term. After that we consider the problem of stability.
Plots of the function $H(\rho)$ (which includes $k_2 H^4$ term)
presented below can be understood in two different ways. First,
the curve $H(\rho)$ can be considered as a track of cosmological 
evolution in some specific regime when $k_3$-term can be neglected.
In this interpretation, $\rho$ is the sum of matter density $\rho_m$ and 
brane tension $\sigma$. The evolution runs from higher to lower energy
(from the right to the left in the plots) till the point $\rho_m=0
, \rho=\sigma$ is reached. On the other
hand, these plots can be interpreted as sets of de Sitter fixed points
(with $\rho=\sigma$ and $\dot H=0$) of a general dynamics with non-zero
$k_3$.

Before we discuss branes it is useful to remember known results
in the standard scenario.
The modification of the standard Friedmann cosmology caused
by the $k_2$-term is shown in Fig.1. The classical cosmology
corresponds to the straight line $H^2 \sim \rho$. The case 
$k_2<0$ results only in changing this dependence  to $H^2 \sim
\sqrt{\rho}$ for large $H$. A positive $k_2$ modifies the situation
significantly. In particular, we can see a new vacuum de Sitter point \cite{m}.
This point is unstable if $k_3<0$, on the other hand, the point $(0,0)$,
representing the late-time Friedmann regime is stable for $k_3<0$. It means
that a trajectory starting in the vicinity of the vacuum de Sitter point has
inflationary behavior at the initial stage, then it leaves a neighborhood
of de Sitter due to instability. As a result, inflation ends, 
and finally the trajectory reaches
the Friedmann late-time attractor. This scenario of acceleration expansion
with the natural exit without any kind of classical  matter is
called as "Starobinsky inflation" \cite{SI}.  

\begin{figure}
\includegraphics[scale=0.4, angle=0]{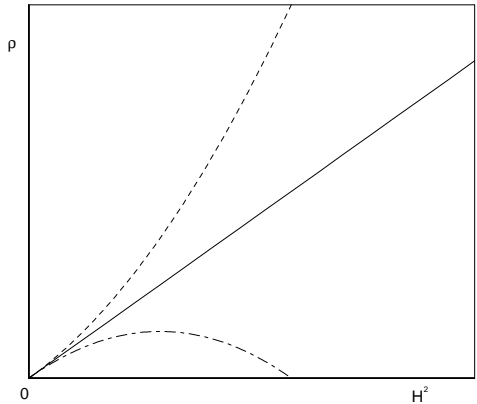}
\caption{The $\rho(H)$ dependence in the standard cosmology without
$k_2$ term of the quantum corrections (solid), with negative $k_2$
(dashed) and with positive $k_2$ (dot-dashed)
}
\end{figure}

We now turn to braneworld models.
In order to understand how $k_2$-term modifies the Randall-Sundrum (RS) brane
it is necessary to remember about two branches, arising
from $\rho^2$ term in the equation of motion. The unmodified equation
for the RS brane is
\begin{equation}
H^2=\Lambda/6+\rho^2/(9M^3).
\end{equation}
Here $\Lambda$ is a cosmological constant in the bulk, which should be
negative in a realistic model, $M$ is the 5-dimensional Planck mass.
The equation (2) when solved with respect to $\rho$ has two solutions
for a given $H$, one with $(+)$-sign and the second with $(-)$-sign before
the square root .   
This equation is represented as a shifted parabola on the plane $(H^2, \rho)$.
As the negative brane tension ultimately leads to
instability, the second branch of the parabola (which corresponds
to negative $\rho$) is unphysical. Positive $k_2$-term changes this
classical picture in a way, very similar to the standard cosmology case
(see the dashed curve in Fig.2) 
with the $(-)$ branch remaining the unphysical one. 
A vacuum de Sitter point now corresponds to a positive root of the forth-order
equation
\begin{equation}
(k_2^2/9M^3) H^8 - H^2 + \Lambda/6=0.
\end{equation}
In the limit $k_2 \to 0$ the value of Hubble parameter in this point tends
to infinity. The existence of a vacuum solution in the RS brane theory with
the quantum corrections was first noticed by Nojiri and Odintsov in
\cite{NO1,NO2}.  
  
The picture for negative $k_2$ is shown in the dot-dashed curve in
Fig.2. The $(+)$ branch 
is similar to the standard cosmology, however, the $(-)$ branch is also
shifted partly to a physical domain $\rho>0$. In particular, it has
a new vacuum de Sitter point. The location the this point is also given
by a positive root of (3), however we will see below that stability properties
of these two de Sitter points are different.

\begin{figure}
\includegraphics[scale=0.4, angle=0]{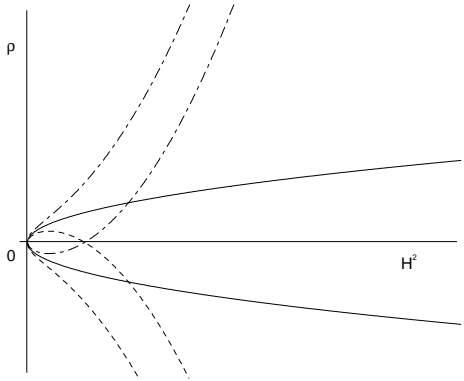}
\caption{The Randall-Sundrum cosmology without $k_2$-term (solid),
with negative $k_2$ (dot-dashed) and with positive $k_2$ (dashed)} 
\end{figure}

Finally, we consider induced gravity (IG) brane. The classical equation of
motion is 
\begin{equation}
m^4 (H^2-\rho/(3m^2))^2 = M^6 (H^2-\Lambda/6).
\end{equation}
Here $m$ is the 4-dimensional Planck mass. 
As the equation of motion is quadratic with respect to $\rho$ 
there are also two branches of solutions.
It should be however noted that in some situations a negative brane tension
on an IG brane does not ultimately lead to instability \cite{Shtanov}, so
negative $\rho$ may have a physical sense in contrast to a RS brane. 
The plots in Fig.3 have been done for the case of Minkowski bulk ($\Lambda
=0$). The $(+)$ branch of the solid curve has a Friedmann-like 
limit when the energy density
of the brane vanishes ($H \to 0$ if $\rho \to 0$), the $(-)$ brane has
a vacuum de Sitter point, founded first by Dvali, Gabadadze and Porrati
in \cite{DGP}. We will call it DGP point.
 
A negative $k_2$ (the dot-dashed curve in Fig.3)
does not change the configurations
of branches. On the other hand, 
positive $k_2$ can modify the picture seriously.
Large enough positive $k_2$ transforms the diagram
so that the lower branch never enters into the $\rho>0$ half-plane,
and the general picture resembles the case of RS brane (compare
the short-dashed curve in Fig.3 with dashed curve in Fig.2). It is evident
that the DGP point is absent in this case. In can be easily calculated
that this situation is realized if $k_2>(4/9)(m^6/M^6)$. 
Nonzero $\Lambda$ in the bulk does not
change this general picture qualitatively,
though the expression for the critical $k_2$ becomes less simple.    
We can see also that the upper branch has a vacuum de Sitter point 
for any positive $k_2$.

\begin{figure}
\includegraphics[scale=0.4, angle=0]{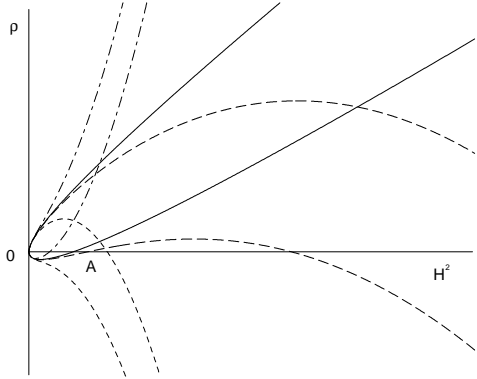}
\caption{The induced gravity brane cosmology without $k_2$-term (solid)
, with negative $k_2$ (dot-dashed), with positive $k_2$ which allows
the DGP point, denoted as $A$ (long-dashed), and with positive $k_2$ 
which does not allow the DGP point (short-dashed)}

\end{figure}

As it was shown recently in \cite{we}, the effective phantom branch
(when $dH/d\rho$ is negative) of IG brane is unstable with respect
to $k_3$-term in the vacuum polarization. Here we will show that this
result is rather general and does not depend on a particular kind of
a brane as well as on the value of $k_2$.

We consider a general situation when matter density in the
corresponding effective Friedmann equation is some
algebraic (may be multi-value) function of the Hubble
parameter $H$, so the evolution equation for a brane has the form:
\begin{equation}
\rho+ \sigma=\tilde F_i(H),
\end{equation}
where the index $i$ marks different branches, $\sigma$ is 
the brane tension.
It is clear that both RS and IG branes
are particular cases of (5).

In the de Sitter point the matter density is diluted, so the equation
becomes

$$
\sigma=\tilde F_i(H).
$$ 

Now we add the quantum corrections (1). 
The $k_2$-term can be incorporated into the algebraic functions
$$
F_i(H)=\tilde F_i(H)+k_2 H^4.
$$

The new de Sitter point is given by

$$
\sigma=F_i(H).
$$

The resulting second order differential equation for $H$
can be written in the form of a system of two first order equations 
$$
\begin{array}{l}
\dot H=C,\\
\dot C=-3CH+\frac{C^2}{2H}+\frac{1}{2Hk_3}(F_{i}-\sigma)
\equiv f_i(H,C).
\end{array}
$$
There is an equilibrium point of this system $(\dot H,\dot
C)=(0,0)$, which corresponds to the de Sitter solution.
In order to
investigate stability of this equilibrium point we need a
linearized system:
$$
\begin{array}{l}
\dot C=(\frac{\partial f}{\partial C})_0C+(\frac{\partial
f}{\partial H})_0H,\\
\dot H=C.
\end{array}
$$

The eigenvalues of this linearized system have the form:
$$
\mu_{1,2}=\frac{1}{2}\left[(\frac{\partial f}{\partial C})_0
\pm\sqrt{(\frac{\partial f}{\partial C})^2_0 +4(\frac{\partial
f}{\partial H})_0} \right].
$$
We can easily see that the first eigenvalue is negative
if and only if
$(\frac{\partial f}{\partial H})_0<0$. The second eigenvalue
is always negative 
(because $(\frac{\partial f}{\partial C})_0=-3H$ is negative
in an expanding universe).

Let us now evaluate $(\frac{\partial f}{\partial H})_0$:
$$
(\frac{\partial f}{\partial
H})_0=\frac{-1}{2H^2k_3}[-\sigma+F_{i}]+\frac{1}{2Hk_3}
\frac{\partial F_{i}}{\partial H}.
$$
In the DeSitter stable point $-\sigma+F_{i}=0$, 
and we have finally
$$
(\frac{\partial f_i}{\partial H})_0=\frac{1}{2Hk_3}(\frac{\partial
F_{i}}{\partial H}).
$$

As a result, if $\frac{\partial F_{i}}{\partial H}>0$
(a normal branch), 
 the de Sitter solution is stable  if $k_3<0$ (as in the standard cosmology).
In the opposite case (a phantom branch) we need $k_3>0$ for stability.

We can conclude, that if we have $k_3<0$ in our Universe (which is
needed for stability of the Minkowski space), all de Sitter 
points on phantom branches are unstable.

Note, that we did not specify any particular modified gravity
theory, which may even be inspired by some  scenario different
from brane models. 
All we need is a function $H(\rho)$. Of course, this analysis can not be 
applied to a situation when equation which relates $H$ and $\rho$
is a differential one (the most important example of this case is 
a scalar field with non-zero potential playing the role of matter). 

It also should be noted, that, strictly spiking, we have proved
the following statement: {\it A de Sitter point, being a future attractor
of a Universe is unstable with respect to vacuum polarization
if it located on a phantom branch}. However, the results of
\cite{we} indicate that an effective phantom regime is
unreachable during a cosmological evolution of a IG brane when
quantum corrections are taken into account, and
it is quite reasonable to suggest that quantum corrections
prevent a classical phantom regime from realization in a general situation.
This problem requires further investigations.

In any cases, we have obtained new restrictions for possible explanations of
a present phantom-like state of "dark energy" which is often claimed
as being favorable by observations \cite{Star,Alcaniz,Paddy}. 
Several years ago it was remarked
that a matter with a constant equation of state parameter $\omega<-1$ causes an
unwanted Big Rip future singularity \cite{S,VS,Caldwell}. This singularity
can be avoided in some scenarios with a phantom scalar field (a scalar field
with the wrong sign of the kinetic term) \cite{Sami, Sami2}, however,
the fact that the energy of a phantom scalar field is unbounded below causes
 severe
problem in quantum theory \cite{Cline} (recently proposed more complicated
models for a scalar field phantom see, for example,
in \cite{Saa,K}). All these problems are absent in modified gravity proposal. 
In this approach the matter
in the Universe remains standard (and, so, there are no problems
with matter instabilities), and
the effective phantom behavior is achieved due to significant modification
of the Friedmann equation. The IG brane is a famous example of
such kind of theory \cite{Dark}. 
Our results, showing that quantum corrections 
have significant influence exactly on 
those branches of cosmological  equations
which simulates a phantom behavior, make clear that the modify gravity
proposal is also not free from instability problems.   
  
Remembering shapes of $H(\rho)$ dependence in braneworld models, we can
see that the new branch of RS brane existing for $k_2<0$ is stable, as
well as new vacuum de Sitter point (3) on this branch. This point has no analog in standard
cosmology and is similar to DGP point on IG brane. On the other hand,
this point in the case of $k_2>0$ is unstable (like in standard cosmology),
giving a possibility of realization of Starobinsky inflation on RS brane.
We also can easily see that the Starobinsky inflation on IG brane is 
always possible (of course, if $k_2>0$) on the $(+)$ branch and for
$k_2<(4/9)(m^6/M^6)$ on the $(-)$ branch.

We have studied modifications of brane cosmology caused by vacuum polarization.
This effect consists of two different terms in an effective energy density.
The term proportional to $H^4$ alters possible fixed points of a cosmological
dynamics, while the term containing time derivatives of $H$ may change
stability properties of these fixed points. A general condition for the future
fixed de Sitter point to be stable have been derived. We should however note
that all these results have been obtained when possible quantum 
corrections in the bulk are neglected. 
This assumption is reasonable in studies of 
quiescent future singularities on a brane, because they are singularities of
embedding \cite{Sht-Sing}, while the bulk remains regular (and, so, far
from a quantum regime). Is quantum corrections in the bulk important
for a brane dynamics in general situation remains unclear,
we leave this problem for future investigations.

\section*{Acknowledgments}

This work is supported by RFBR grant 05-02-17450
and scientific school grant 2338.2003.2 of the Russian Ministry
of Science and Technology. Authors are greatful to Alexey Starobinsky,
Varun Sahni and Yuri Shtanov for discussions.

\end{document}